\documentclass[aip, apl, reprint,floatfix,superscriptaddress]{revtex4-2}
\usepackage[utf8]{inputenc}
\usepackage{amsmath}
\usepackage{dcolumn}
\usepackage{natbib}
\usepackage{soul}
\usepackage{ulem}
\usepackage{graphicx}
\usepackage{mathrsfs}
\usepackage{bm}
\makeatletter
\newcommand{\RNum}[1]{\uppercase\expandafter{\romannumeral #1\relax}}
\usepackage{float}
\makeatother
\usepackage{color,soul}
\usepackage{amssymb}
\usepackage{gensymb}

\begin{document}
\title{Electric field control of RKKY coupling through solid-state ionics}

\author{Maria Ameziane}
\affiliation{NanoSpin, Department of Applied Physics, Aalto University School of Science, P.O. Box 15100, FI-00076 Aalto, Finland} 

\author{Roy Rosenkamp}
\affiliation{NanoSpin, Department of Applied Physics, Aalto University School of Science, P.O. Box 15100, FI-00076 Aalto, Finland} 

\author{Luk\'{a}\v{s} Flaj\v{s}man}
\affiliation{NanoSpin, Department of Applied Physics, Aalto University School of Science, P.O. Box 15100, FI-00076 Aalto, Finland} 

\author{Sebastiaan van Dijken}
\affiliation{NanoSpin, Department of Applied Physics, Aalto University School of Science, P.O. Box 15100, FI-00076 Aalto, Finland}

\author{Rhodri Mansell}
\email{rhodri.mansell@aalto.fi}
\affiliation{NanoSpin, Department of Applied Physics, Aalto University School of Science, P.O. Box 15100, FI-00076 Aalto, Finland}

\begin{abstract}
Placing a suitable spacer layer between two magnetic layers can lead to an interaction between the magnetic layers known as  Ruderman–Kittel–Kasuya–Yosida (RKKY) coupling. Controlling RKKY coupling, particularly the ability to switch between ferromagnetic and antiferromagnetic coupling, would enable novel magnetic data storage devices. By combining solid-state Li ion battery technology with an out-of-plane magnetized Co/Pt-based stack coupled through a Ru interlayer we investigate the effects of the insertion of Li ions on the magnetic properties of the stack. The RKKY coupling and its voltage dependence is measured as a function of the Ru interlayer thickness, along with the effects of repeated voltage cycling. The Li ions both change the amplitude of the RKKY coupling and its phase, leading to the ability to switch the RKKY coupling between ferromagnetic and antiferromagnetic with applied voltages. 

\end{abstract}

\maketitle
The ability to control magnetism through applied voltages opens a path to low energy magnetic data storage devices \cite{Matsukura_2015,Song_2017,Liang_2021}. Among the various approaches to using voltages to control magnetism, magneto-ionics has recently seen increased interest due to the large effects obtainable with this approach \cite{Nichterwitz_2021,Gu_2021,deRojas_2022}. The insertion of non-magnetic ions into magnetic layers has been shown to change important magnetic properties such as the saturation magnetization \cite{bi2014reversible,bauer2015magneto,tan2019magneto}, magnetic anisotropy \cite{bauer2015magneto,tan2019magneto,Ameziane_Li_PMA_AdvFuncMat_2022}, Dzyaloshinskii-Moriya interaction \cite{srivastava2018large,HerreraDiez_IonicDMI_PRA_2019} as well as exchange bias \cite{Gilbert_ExchangeBiasRedox_NatCOmms_2016} and ferrimagnetic order \cite{Huang_Ferri_H_ion_NatNano_2021}. However, for applications in digital memory and logic it is preferable not to change the magnitude of a magnetic property but to cause 180$^\circ$ switching. This is not straightforwardly achieved with electric fields, which lack the time-symmetry breaking property of magnetic fields \cite{Huang_Ferri_H_ion_NatNano_2021}. Whilst pulse switching is possible \cite{Shiota_2012}, this requires finely tuned magnetic parameters. One path is to use two magnetic layers coupled through RKKY interactions \cite{ParkinCoRuPhysRevLett.64.2304,PhysRevB.44.7131,PhysRevB.46.261,Duine_SAF_review_NatPhys_2018,Kossak_RKKY_Voltage_2023}. The coupling derives from spin-dependent reflection of the electron wavefunction at the normal metal / magnetic metal interface. This leads to a coupling that oscillates between antiferromagnetic and ferromagnetic coupling as a function of the thickness of the normal metal spacer layer, which is characterized by the wavelength and phase of the oscillation as well as the decay length of the envelope of the oscillation \cite{ParkinCoRuPhysRevLett.64.2304,PhysRevB.46.261}. 

RKKY coupling is sensitive to changes to the system, as it depends on the electrons at the Fermi surface of the interlayer \cite{PhysRevB.46.261}. This means the coupling can be modified for instance by doping the interlayer \cite{Bruno_Interlayer_Disorder_JMMM_1997}, modifying the capping layer \cite{PhysRevB.56.8919}, or changing the band filling in the spacer layer \cite{PhysRevB.100.014403}. Control of RKKY coupling is a promising target for devices that aim to control magnetism through electric fields. As well as theoretical proposals for devices \cite{801049}, control has been demonstrated in several experimental systems based on liquid ion gating \cite{Yang_RKKY_IonicGating_NatCOmms_2018} and voltage-induced switching in magnetic tunnel junctions \cite{Newhouse_EField_MTJ_NatComms_2017,Zhang_MTJ_NanoLett_2022}. However, the switching of tunnel junctions still involves significant current densities. Switching of magnetic layers through voltage control of RKKY interactions has been achieved in Co-based perpendicularly magnetized layers using the insertion of hydrogen ions \cite{Kossak_RKKY_Voltage_2023}.  


Here we investigate the electric field control of RKKY coupling using a solid-state Li ion based device incorporating a Li storage layer, lithium cobalt oxide (LCO), and an ionic conductor, lithium phosphorous oxynitride (LiPON). By using technology taken from the field of solid-state Li ion batteries a large density of Li ions can be provided at low voltages \cite{Ameziane_Li_PMA_AdvFuncMat_2022}. Perpendicularly magnetized Co/Pt layers are used for the fixed and free layers of the device which are coupled through a wedged Ru layer. Applying a positive voltage to the top electrode of a junction
causes Li ions to move from the storage layer through the ionic conductor to the top layers of the metallic stack \cite{Ameziane_Li_PMA_AdvFuncMat_2022}. 

\begin{figure}[htb]
\includegraphics[width=1.0\linewidth]{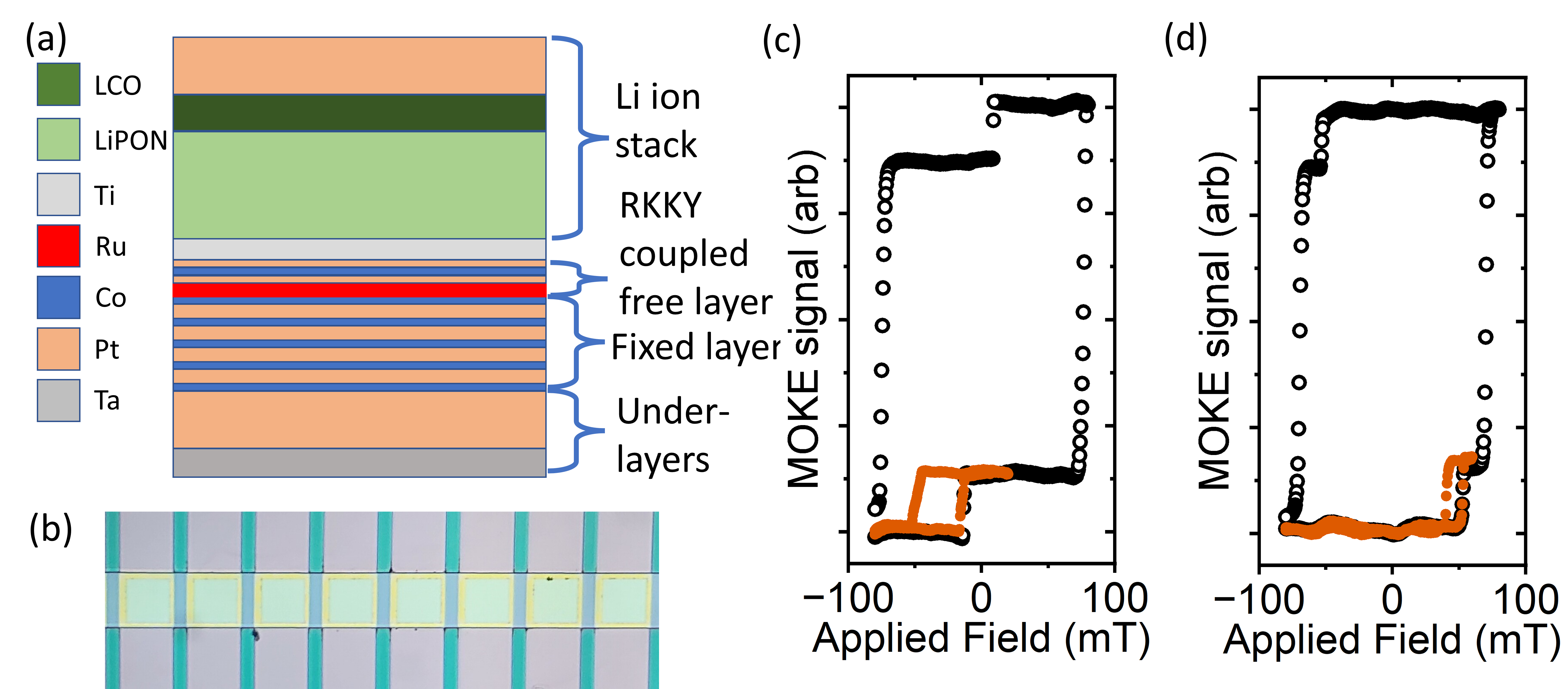}

\caption{(a) Cross-sectional schematic of a junction. (b) Optical microscopy image of the device consisting of vertical bottom electrodes and a horizontal top electrode. The bottom electrodes are 100 $\mu$m across. The thickness of the Ru interlayer increases from left to right.  (c) Major hysteresis loop (black) and minor loop (orange) of the junction with 1.0 nm Ru at 0 V showing antiferromagnetic coupling between the free and fixed magnetic layer. (d) Major hysteresis loop (black) and minor loop (orange) of the junction with 2.4 nm Ru at 0 V showing ferromagnetic coupling.}
\label{Fig1}
\end{figure}

In Fig. 1(a) we show a cross-sectional schematic of a junction. The total structure consists of a metal bottom electrode which contains all of the magnetic layers. The metallic stack is Ta (2 nm) / Pt (4 nm) / [Co (1 nm) / Pt (1 nm)]$_4$ / Co (1 nm) / Ru wedge / Pt (0.25 nm) / Co (0.4 nm) / Pt (0.25 nm) / Ti (1.5 nm). The bottom Co/Pt multilayer below the Ru wedge acts as the fixed layer of the device, with the top Co single layer acting as the free layer. The 0.25 nm Pt layers around the top Co layer lead to perpendicular magnetization while still preserving the RKKY coupling \cite{LavrijsenPMA_RKKY_APL_2012}. The upper layers form the Li ion conduction and storage part of the device and consist of LiPON (70 nm) / LCO (20 nm) / Pt (5 nm). The bottom electrode, consisting of the metallic layers capped by Ti, is patterned by optical lithography and lift off into 100 $\mu$m wide stripes, where the length of the stripes is orthogonal to the direction of the Ru wedge. A second lithography step is used to create an insulating SiN layer with windows. The top electrode then creates cross junctions with the metallic bottom electrode through the windows in the SiN layer. The SiN acts to reduce shorting at the edges of the junctions. The Ru wedge thickness is estimated from a calibration sample grown with the same wedge parameters as the device. 

The fabrication process results in an array of crossbar junctions each with a different average thickness of the Ru interlayer. Figure 1(b) shows an optical microscopy image of the device structure. The change of the Ru thickness within each junction is of the order of 0.3 $\textrm{\AA}$ and the thickness increases from left to right. . The bottom electrode is grounded and voltages are applied to the top electrode using a Keithley sourcemeter. To demonstrate the magnetic behavior of the junctions we show in Fig. 1(c) the junction with a  Ru thickness of 1.9 nm. The major hysteresis loop (black) shows two switches coming from negative saturation, firstly the smaller switch of the top Co single layer before 0 mT, followed by the switch of the Co/Pt multilayer at around +80 mT. The thin Pt layers lead to exchange coupling between the bottom five Co layers so that they switch as an effective single layer. The top Co layer is RKKY coupled to the bottom layers. This leads to an effective bias field on the layer, which itself depends on the thickness of the Ru interlayer. Here, the top layer minor loop (orange) shows antiferromagnetic coupling. Coming from negative saturation the top layer switches already at negative fields so that the top layer is aligned antiparallel to the bottom layers at zero magnetic field. In Fig. 1(d), at a different point on the wedge with 2.4 nm of Ru, we measure a minor loop that corresponds to ferromagnetic coupling. Here the minor loop is shifted to positive applied fields, showing that the RKKY coupling favors parallel alignment of the layers. The RKKY coupling also affects the coercive field of the bottom layers, seen in their reduced coercivity in Fig. 1(d) compared to Fig. 1(c), but the effect is small due to the greater combined moment of the bottom layers.

\begin{figure}[bht]
\includegraphics[width=1.0\linewidth]{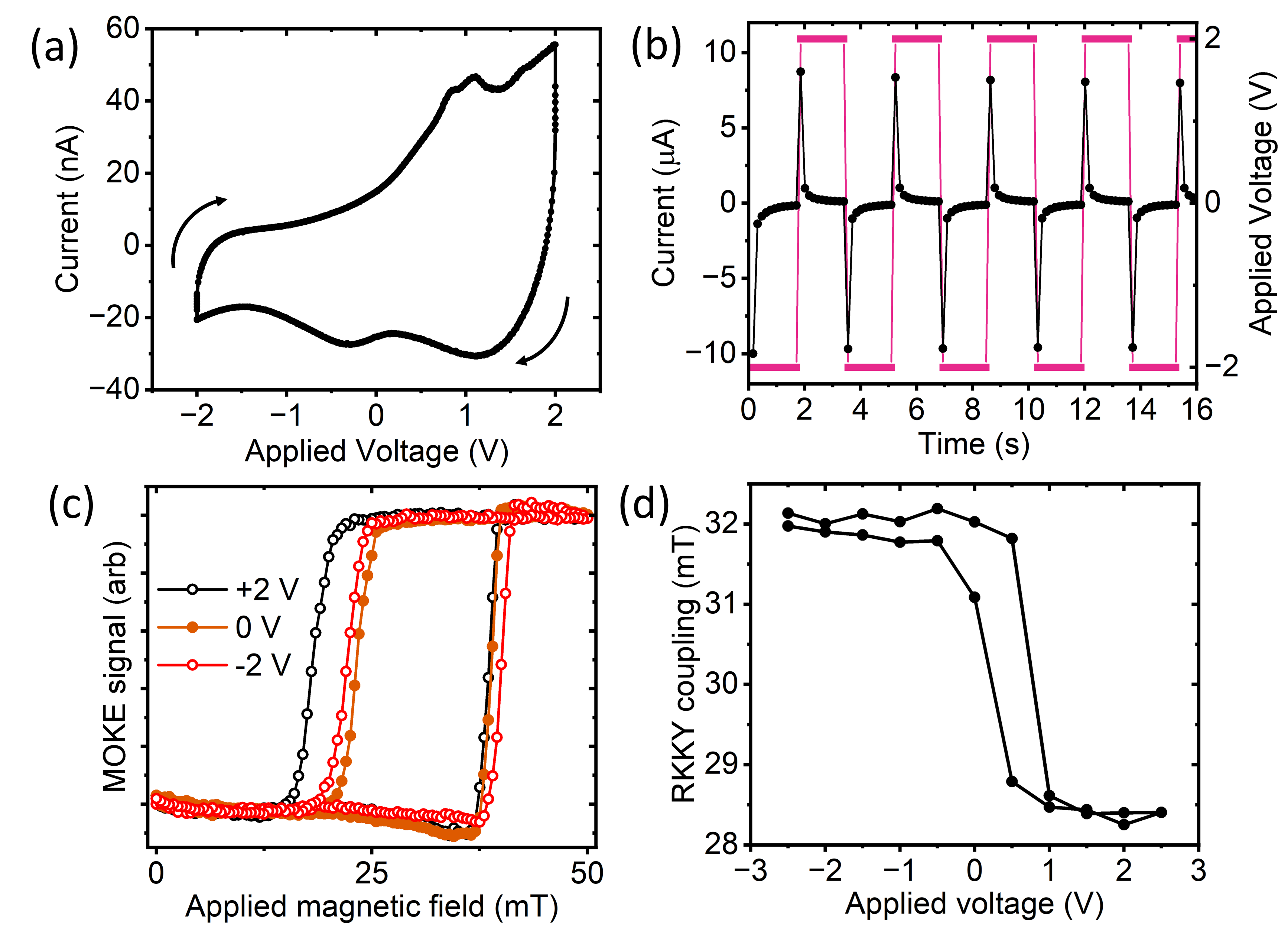}

\caption{(a) Cyclic voltammagram of the junction with 1.9 nm Ru interlayer taken at 50 mV/s. (b) Current flow (left axis) through the same junction as (a) driven by $\pm$ 2 V toggle switching (right axis). (c) Hysteresis loops of the junction with 2.6 nm Ru recorded at three different applied voltages. (d) The RKKY coupling of the 2.6 nm Ru junction when cycling the voltage from $-2.5$ V to +2.5 V and back to $-2.5$ V. At each voltage a minor loop is taken to determine the RKKY coupling strength.}
\label{Fig2}
\end{figure}

The junctions also have significant electrical properties, derived from their battery-like structure. The cyclic voltammagram of the junction with 1.9 nm Ru interlayer thickness is shown in Fig 2(a). The measurement shows an asymmetric loop with notable redox peaks at around +1 V on the positive sweep and a broader peak at $-0.5$ V on the downward sweep. The asymmetric loop shown here is typical of an intercalation dominated electrochemical process \cite{Simon_Gogotsi_ElectroChem_Review_2020}. All the junctions measured show similar behavior. We also cycled the junctions with a stepped voltage as shown in Fig. 2(b). A stepped voltage is more technologically relevant than the slow sweep of the cyclic voltametry. Stepping the voltage between -2 V and +2 V leads to peak currents larger by two orders of magnitude passing through the devices. These currents however, also quickly decay with a time constant under a second, showing the relatively rapid ion movement possible with Li ion-based devices. In Fig. 2(c) we show how applied dc voltages effect the magnetic layers. The minor hysteresis loops from a junction with 2.6 nm Ru are shown, which shows ferromagnetic coupling (see also Fig. 1(d)). At negative voltages a relatively narrow loop is seen with an offset of around +32 mT, which remains at 0 V. At +2 V applied, which corresponds to the insertion of Li ions into the magnetic layers, the magnitude of the bias decreases to around 27 mT and the hysteresis loop broadens. In Fig. 2(d) we show the RKKY coupling strength of this junction as the voltage is cycled from $-2.5$ V to +2.5 V and back in steps of 0.5 V. The RKKY coupling shows hysteretic behavior, with the switching occurring around +1 V in the positive sweep direction and 0 V in the negative sweep direction, roughly consistent with the peaks seen in the cyclic voltammagram (Fig. 2(a)). 

\begin{figure}[thb]
\includegraphics[width=1.0\linewidth]{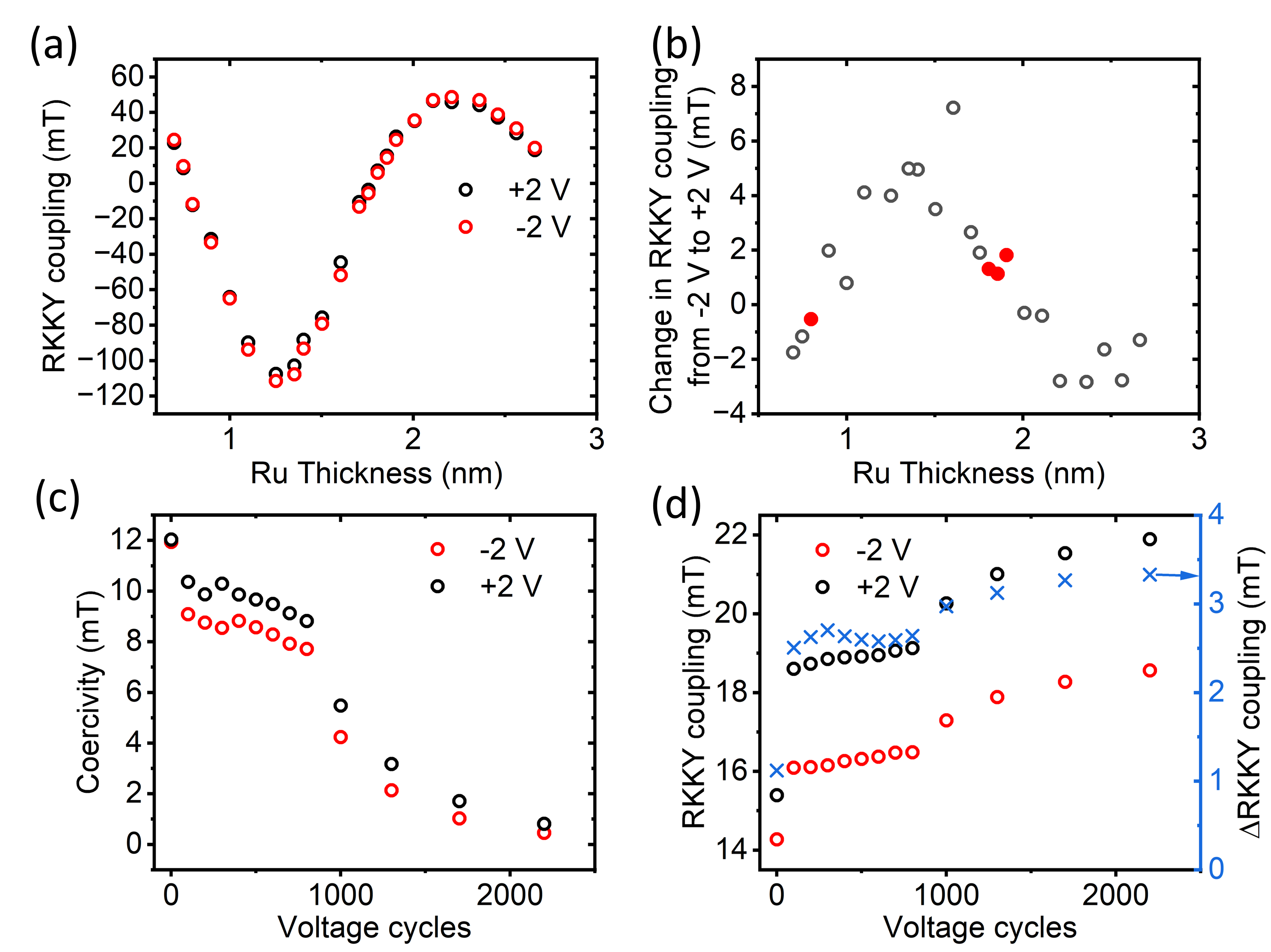}

\caption{(a) RKKY coupling strength as a function of Ru interlayer thickness for +2 V and $-2$ V. (b) Difference in RKKY coupling strength for +2 V and $-2$ V as a function of Ru interlayer thickness. (c) Coercivity of the junction with 1.85 nm Ru interlayer thickness at +2 V and $-2$ V as a function of voltage cycles. The cycling was carried via $\pm$ 2 V switching as in Fig. 2(b). RKKY coupling of the same junction as (c) at +2 V and $-2$ V as a function of voltage cycles (left axis). The voltage induced change in RKKY as a function of voltage cycles (right axis).}
\label{Fig3}
\end{figure}

Combined voltage-dependent RKKY coupling data for all the junctions measured is shown in Fig. 3(a) as a function of the Ru interlayer thickness. The RKKY coupling is shown for +2 V and $-2$ V where the effect of voltage is small compared the the effect of the changing Ru thickness. Negative values are used to indicate antiferromagnetic coupling, with positive values corresponding to ferromagnetic coupling across the Ru interlayer. As a function of interlayer thickness a peak in antiferromagnetic RKKY coupling is seen around 1.2 nm Ru followed by a ferromagnetic peak at around 2.1 nm. This is similar to what is expected from previously studied Co/Ru coupling systems, although the peak antiferromagnetic coupling is measured at slightly larger Ru thickness \cite{ParkinCoRuPhysRevLett.64.2304,LavrijsenPMA_RKKY_APL_2012}. In Fig. 3(b) the difference between the $\pm$ 2 V data is plotted. Generally, the effect of positive voltages, which cause the insertion of Li ions into the magnetic layers, is to reduce the magnitude of the coupling for both antiferromagnetically coupled and ferromagnetically coupled junctions. However, there is a further effect. The symbols plotted in red show where the effect is reversed, that is, positive voltages cause an increase in the strength of RKKY coupling. This effect occurs around the crossings between antiferromagnetic and ferromagnetic coupling and indicates that as well as a change in strength there is also a change in the phase of the RKKY coupling caused by the insertion of Li ions. 

\begin{figure}[htb]
\includegraphics[width=1.0\linewidth]{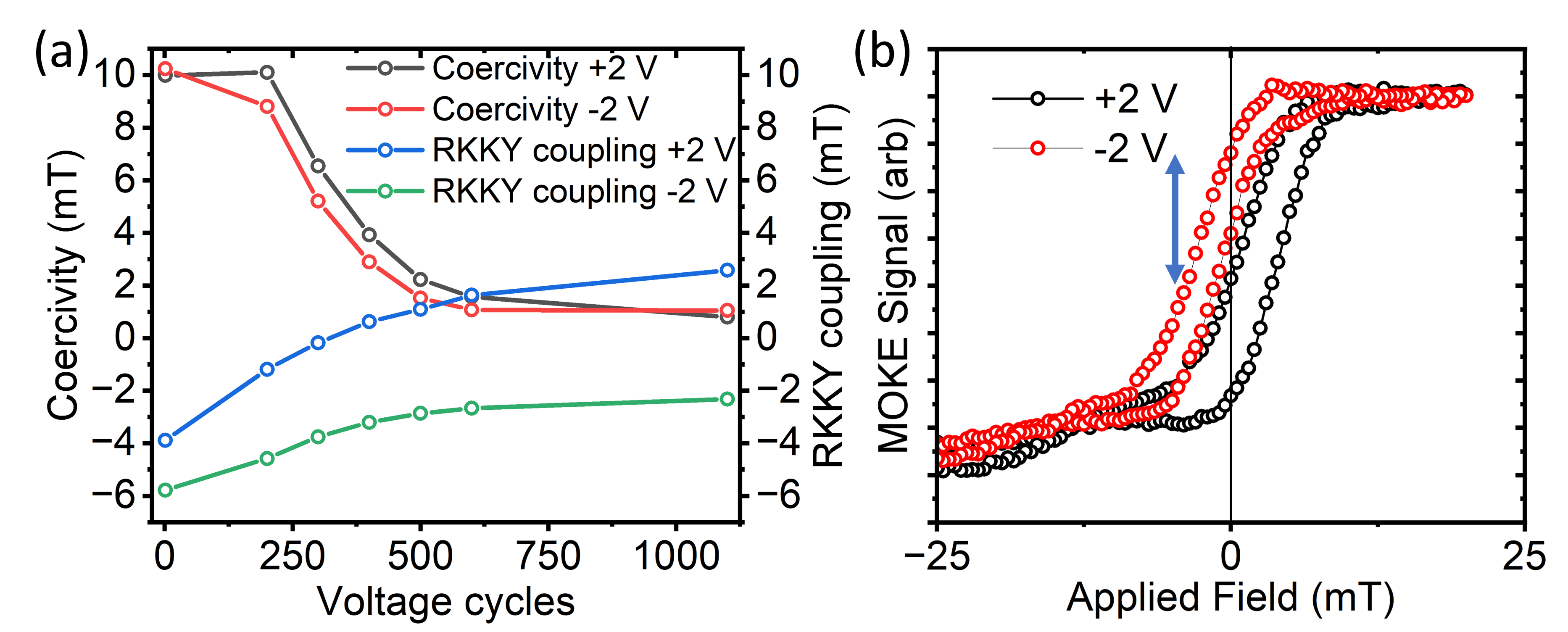}

\caption{(a) Changes in the coercivity and RKKY coupling of the junction with 1.75 nm Ru at $-2$ V and +2 V as a function of the number of voltages cycles. (b) Minor loops at $-2$ V and + 2 V after 600 voltage cycles. The blue arrow shows the extent of all-electrical zero-field switching determined from minor loops taken after electrical switching.}
\label{Fig4}
\end{figure}

In Fig. 3(c)-(d) we show the effect of extensive voltage cycling on the magnetic properties of the junctions. The junction with 1.85 nm Ru is cycled as in Fig 2(b) using $\pm$ 2 V square pulses. In Fig. 3(c) the change of coercivity is shown as a function of the number of cycles, demonstrating a significant drop from an initial coercivity of 10 mT down to less than 1 mT after 2000 cycles. At the same time, as shown in Fig. 3(d), the RKKY coupling becomes more positive, that is its magnitude increases for both $-2$ V and +2 V. This leads to an increase also in the size of the voltage effect on the RKKY coupling from around 1 mT initially to around 3 mT after extensive cycling. This effect is most likely caused by changes to the top Co single layer. The cycling of Li ions may disrupt the Co/Pt interfaces leading to lower anisotropy and lower effective thickness of the Co layer. The lower anisotropy is likely to lead to lower coercivity, whilst a decreased effective thickness of the Co layer will lead to a higher effective RKKY coupling field.

The effects shown in Fig. 3 can be used to create zero magnetic field switching of magnetization under an applied voltage. In Fig 4(a) we show the effect of cycling the junction with 1.75 nm Ru interlayer thickness which is slightly ferromagnetically coupled before the application of voltages. The initial coercivity of the layer is around 10 mT and this is significantly larger than the effects of voltage on the RKKY coupling ($\sim$ 2 mT). By cycling the junction the coercivity is reduced and the RKKY coupling at the different voltages shifts. After 600 cycles the coercivity has dropped below 2 mT, the RKKY coupling at +2 V has become positive, whilst the RKKY coupling at $-2$ V is still negative. Firstly, this demonstrates clearly the effect of the Li ions on the phase of the RKKY coupling, as the coupling can be switched from ferromagnetic to antiferromagnetic by the applied voltage. This is the necessary condition to creating devices based on voltage control of the RKKY coupling. Secondly, the shift of the RKKY coupling caused by the voltage is larger than the coercivity, and so it should be possible to switch the magnetization at zero magnetic field. In Fig. 4(b) the minor hysteresis loops after 1100 cycles are shown. Although the shift in the RKKY coupling is greater than the coercivity, the loops are slanted, which is consistent with a reduced anisotropy. From loops starting at zero applied field taken after electrical cycling the extent of the all-electrical switching is shown by the blue arrow in Fig. 4(b) and is equal to around a third of the total magnetization. The effect of the electrical switching is to shift the net moment between the zero magnetic field positions on the same side of the hysteresis loop, which is determined by the initialization, rather than to cause a crossing of the loop. 

In summary, we have shown that the insertion of Li ions under applied voltages can be used to control the strength of RKKY coupling in a system with perpendicular magnetization. The effect of the Li ions is not just to alter the amplitude of the RKKY coupling, but also its phase. We have shown that RKKY coupling can be tuned between antiferromagnetic and ferromagnetic with an applied voltage. We were then able to demonstrate partial switching of magnetization under an applied voltage. The results suggests that magneto-ionic control of RKKY coupling is a promising approach for the creation of fully voltage switched magnetic memory devices.

\section*{Acknowledgments}
This work was funded by the Academy of Finland under project numbers 316857 and 295269. We acknowledge the provision of facilities by Aalto University at OtaNano - Micronova Nanofabrication Centre.

\bibliography{ref,reference}

\end{document}